\documentclass{llncs}
 
\usepackage[utf8]{inputenc}
\usepackage{eurosym}

\usepackage{graphicx}
\usepackage{float}

\usepackage[fleqn]{amsmath}
\usepackage{tabularx}
\usepackage{tabulary}
\usepackage{mathpartir}
\usepackage{amssymb}

\usepackage{color}
\usepackage{xcolor}
\usepackage{framed}
\definecolor{shadecolor}{rgb}{0.9,0.9,0.9}
\definecolor{Orange}{rgb}{1,0.5,0}

\usepackage[pdftex]{hyperref}
\hypersetup{%
pdfauthor={blind submission}, %
bookmarksnumbered, %
pdfstartview={c}, %
colorlinks,%
citecolor=black, %
filecolor=black, %
linkcolor=black, %
urlcolor=black}

\usepackage{mathtools}
\usepackage{cite}
\usepackage{epsfig}
\usepackage{epstopdf}
\usepackage{algorithm}
\usepackage{algorithmic}
\usepackage{caption}

\newcommand{\provGen}{\textit{provGen}}
\newcommand{\used}{\mathit{used}}

\floatname{algorithm}{Procedure}

\title{ProvGen: generating synthetic PROV graphs  with predictable structure}

\author{Hugo Firth \and Paolo Missier}

\institute{School of Computing Science, Newcastle University, UK\\
           \{h.firth, paolo.missier\}@ncl.ac.uk
}

\begin{document}
\maketitle

\begin{abstract}
This paper introduces \provGen{}, a generator aimed at producing large synthetic provenance graphs with predictable properties and of arbitrary size.
Synthetic provenance graphs serve two main purposes. 
Firstly, they provide a variety of controlled workloads that can be used to test storage and query capabilities of provenance management systems at scale. 
Secondly, they provide challenging testbeds for experimenting with graph algorithms for provenance analytics, an area of increasing research interest.
\provGen{} produces PROV graphs and stores them in a graph DBMS (Neo4J). 
A key feature is to let users control the relationship makeup and topological features of the graph, by providing a seed provenance pattern along with a set of constraints, expressed using a custom Domain Specific Language. 
We also propose a simple method for evaluating the quality of the generated graphs, by measuring how realistically they simulate the structure of real-world patterns.
\end{abstract}


\section{Introduction} \label{section:introduction}

Every piece of data ever produced, either manually or automatically, has a provenance. 
This is metadata that provides an account of how the data was created.
Examples include a blog's author, the history of a piece of software along with its contributors, the instruments used to take a measurement, and their settings; or a description of an experimental process used to produce a scientific result.
%
%
%
The PROV data model for provenance~\cite{w3c-prov-dm}, endorsed in 2013 by the W3C, provides a formal and domain-agnostic grounding for provenance, in the form of UML and OWL models, and RDF, XML, and relational (PROV-N\cite{w3c-prov-n})
serializations.
We refer to PROV instances as digraphs, where nodes are of three possible types: \textit{Entities} (for data, documents, anything that has provenance), \textit{Activities}, which model the execution of a data consumption and production process; and \textit{Agents}, to whom Entities can be attributed, and who hold responsibility for carrying out Activities.
The edges represent instances of relationships amongst the nodes, which are documented in the PROV-DM specification~\cite{w3c-prov-dm}.

\begin{figure*}[htb]
\begin{center}
\includegraphics[width=.6\textwidth]{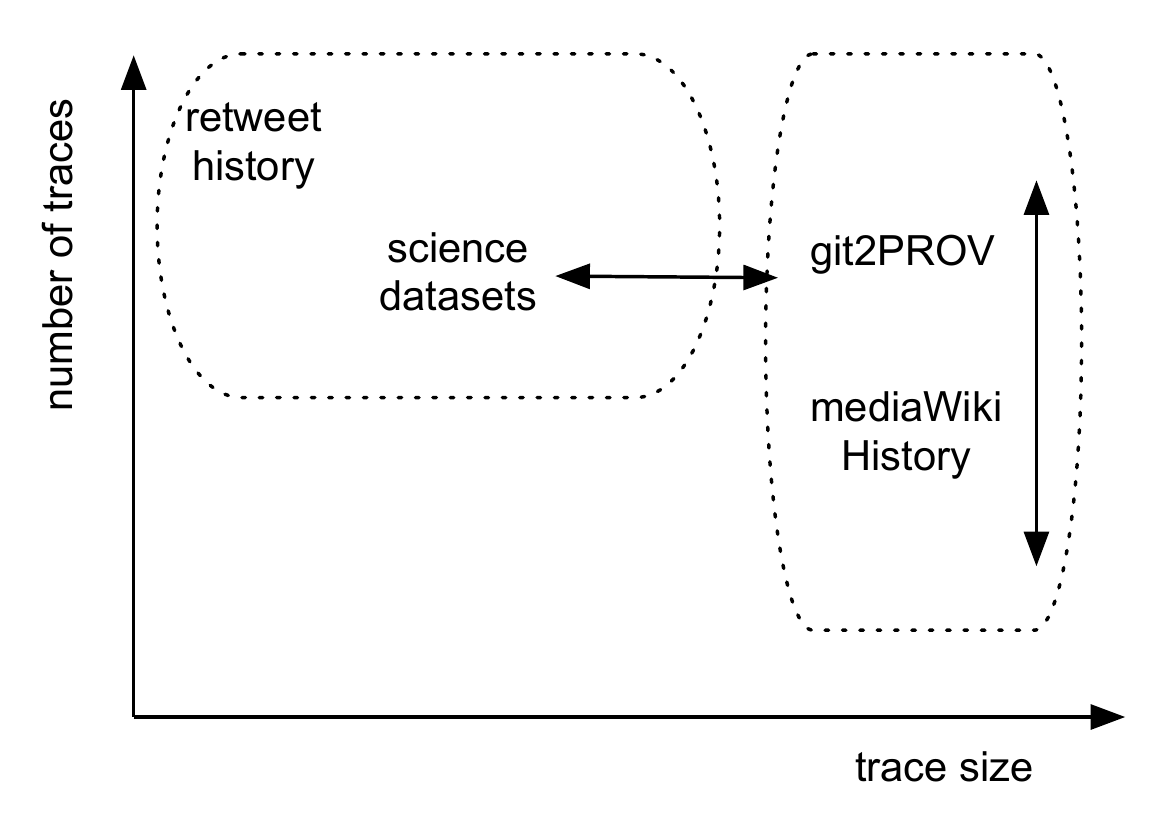}
\caption{A simple space for homogeneous provenance collections}
\label{fig:prov-quadrant}
\end{center}
\end{figure*}

The provenance traces associated with a homogeneous data collection (a scientific data repository, all the blogs hosted on a particular site, all the artifacts associated with a complex software project) also naturally form a collection.
Such collections grow in size both with the number of underlying data products, and with the complexity of their production process.
Fig.\ref{fig:prov-quadrant} suggests how different collections can be placed into a space defined by volume, i.e., the number of traces in a collection, and by the typical size of a trace within a collection.
For instance, many small traces (upper left) may be associated with a large repository of scientific data, while complex software with a long history may be represented by many large traces (upper right), as exemplified by the Git2Prov\cite{DBLP:conf/semweb/NiesMVCGMW13} tool.
%

Arguably, the value of provenance comes not only from querying the content of individual traces, but also from analytics, which can only be computed on whole collections.
It is therefore important for practical applications to demonstrate the effectiveness of a data and service architecture to manage large bodies of provenance, with special focus on the upper quadrant of our size/volume space.
Thus, we expect that the design of scalable repositories for provenance traces should be a natural concern in provenance management. 
A number of recent efforts have been documented on nascent provenance management infrastructure \cite{Chebotko,chebotko2010rdfprov,6009254,CPE:CPE1870}, and there is evidence of the emergence of applications that require provenance querying in a variety of settings (eg \cite{Mattoso:2013:UHW:2499896.2499900,deusing}).
However, unlike other ``big data'' domains such as Linked Data and more generally RDF triple stores, where performance benchmarking is established practice, to the best of our knowledge no community-made benchmarking and commonly accepted datasets that are specific to provenance are available.\footnote{The use of community datasets for comparing the performance of predictive models has also long been commonplace within the data mining and KDD community, where challenge datasets are regularly used.}
This makes it difficult to benchmark and compare different implementations with regards to storage techniques, query models, and analysis algorithms.

This is somewhat counter-intuitive, given the amount of provenance that is generated, in domains such as those alluded to above.
In fact, only a handful of real datasets are currently available through a community process, i.e., the first ProvBench initiative in 2013 (\url{http://bit.ly/1fBOswR})\footnote{Further contributions are expected from the second ProvBench in 2014 (\url{http://bit.ly/1c0q5rS}).}, and even fewer conform to the recent PROV standard and are therefore interoperable.
Existing benchmarking datasets which apply to RDF triple stores\footnote{The W3C maintains a list of those \url;{http://www.w3.org/wiki/RdfStoreBenchmarking}} are not adequate, because they fail to account for the specific data model and semantics of PROV, as well as for the specific requirements of provenance query and analysis. 

\subsection{Contributions} 

Our assumption is that synthetic PROV graphs can be a valuable complement to emerging natural provenance collections, provided that their structural properties reflect specific provenance patterns, with control over their repetition and variability, and at varying scales.
Such graphs can be used both for benchmarking emerging provenance management systems, as well as to test analytics algorithms that operate naturally on large provenance collections.

\begin{figure}
  \includegraphics[width=\textwidth]{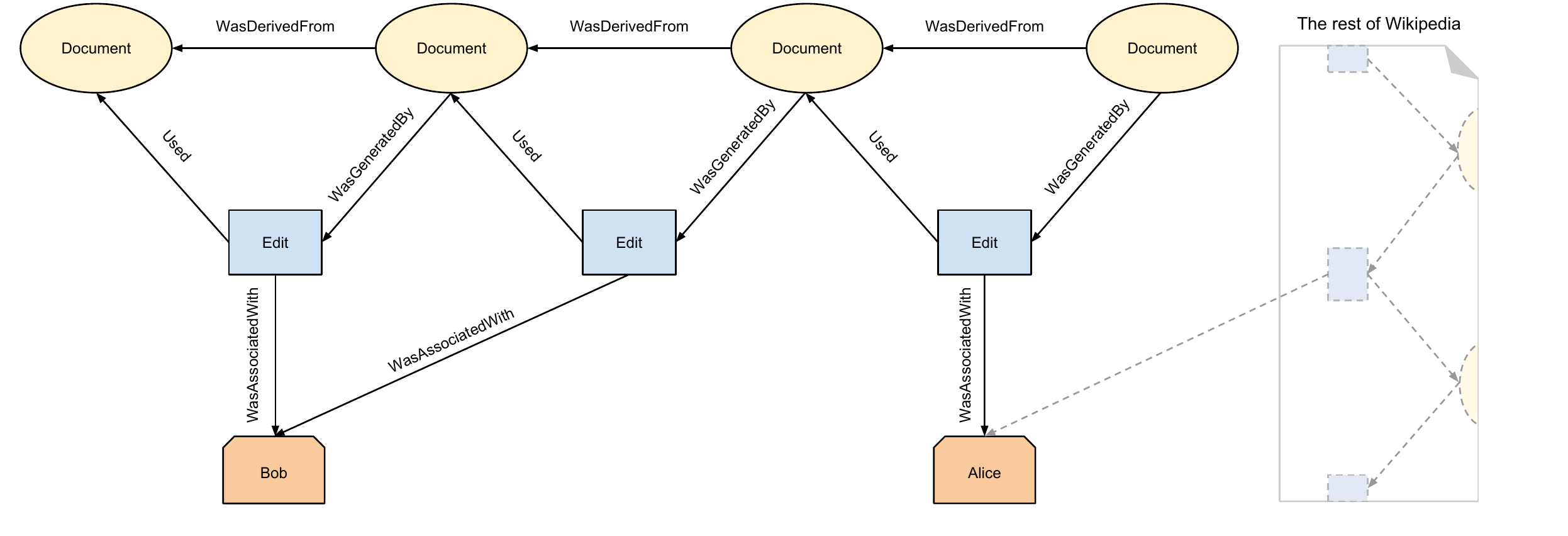}
  \caption{The \textit{document revision} provenance pattern in Wikipedia includes multiple derivation and editing activities by multiple user or bot agents.}
  \label{figure:wikipedia-pattern}
\end{figure}

Our main  contribution (Sec. \ref{section:prov-gen-model}) is the design and implementation of \provGen, a PROV generator that is designed to help populate the space described in Fig.\ref{fig:prov-quadrant}.
\provGen{}  ``grows'' collections of synthetic PROV graphs in a way that conforms to real-life provenance patterns. These are currently user-defined and modelled after patterns found in specific domains, and which reflect the nature of the data generation process described by the provenance.
For instance, the prevalent provenance pattern for  a Media Wiki website, which we refer to as the ``document revision" model, involves multiple revisions of articles, by multiple editors (Fig.\ref{figure:wikipedia-pattern}).
Git repositories exhibit similar patterns, which reflect the revision history of the code. 
These patterns are different, for instance, from those for the provenance of data generated using a workflow, which reflect the consumer-producer graph structure of the dataflow specification.

Users control the ``shape'' of the graph being generated by \provGen{} by providing two main elements.
The first is a  \textit{seed graph}, which determines the specific types of nodes and the relationships amongst them to be considered, in an otherwise random generation process.
 The second element is a set of constraints, expressed using a dedicated Domain Specific Language (DSL), which limit the possible ways in which nodes and relationships are added.
These two elements ensure a predictable general shape for the generated graph, as well as its compliance to PROV.

\provGen{} relies on a graph DBMS back (Neo4J). In particular, the generation algorithm is based on graph rewrite rules that are implemented using a combination of Cypher queries and Create statements.
    
\subsection{Related work} \label{section:related-work}

%

A growing body of research is devoted  to generating large bodies of synthetic graph data, either using purely random models \cite{Karrer2009, Erdos1960}, or by generating graphs that exhibit specific statistical properties\cite{Barabasi1999, Batagelj2005, Leskovec2005}. 
One example is the \textit{preferential attachment} model. Popularised by Barabasi and Albert\cite{Barabasi1999}, this model states that as new vertices are added to a graph, the probability of creating a relationship with node \textit{n} is inversely proportional to the degree of \textit{n}. This model generates a graph with a degree distribution which follows a power law. 

An issue common to these models, emphasised for instance in a comprehensive survey on graph generators \cite{Chakrabarti2006}, is their focus on enforcing global properties of the generated graph, such as degree distribution, clustering coefficient, etc. 
A potential reason for this focus is that these generators are aimed at simulating social networks \cite{Pham2013, Batagelj2005}, the statistical properties of which are based on large sets of examples, and thus are fairly well understood \cite{Mislove2007}. 
In contrast, our generation strategy relies on user-specified patterns, rather than a large set of pre-existing examples (in the future, we hope to be able to use patterns that have been automatically discovered from existing graphs, by means of standard graph mining techniques \cite{Kuramochi2004}).
This has the advantage that the overall topology of the graph can be made to reflect desired semantic properties of the data, such as the average number of usages for a certain type of entity, the average number of association of an agent with activities, and so forth.
Pham \textit{et al}.\cite{Pham2013} are amongst the few to have addressed this problem. However, they focus on a loosely related issue, namely the correlation between node and relationship properties, such as an increased likelihood to be called ``Joachim'' if you live in Germany, and on generating realistic synthetic value dictionaries accordingly.

\section{Graph generation model} \label{section:prov-gen-model}

Graph generation in \provGen{} is an iterative process which starts from a single node.
At each iteration, a collection of predefined \textit{atomic rewrite} rules is used to add a set of new nodes or relationships to the current graph.
These rules account for all possible relation types that are defined in the PROV-DM specification.
As an example, consider the definition of the $\used(a,e)$ relation between an activity $a$ and $e$ an entity $e$.
Three atomic graph rewrite rules are defined for this relation, namely (i) given an entity node $e$, add a new activity node $a$ and an edge $\used(a,e)$; (ii) given an activity node $a$, add a new entity node $e$ and an edge $\used(a,e)$; and (iii) given a pair of unrelated nodes $(a,e)$, add edge $\used(a,e)$.
Since each single PROV relation type induces three atomic rewrites, and we consider 13 types of relations from PROV, at each iteration \provGen{} can potentially fire any of 39 different rules.

Users can control the execution of these rules and the overall effect of the generation process in three complementary ways, namely (i) by specifying a \textit{seed graph}, (ii) by adding a set of constraints, and (iii)  by specifying additional execution parameters.
We now describe these in some detail.

\paragraph{1. Seed graphs.} 
A seed graph specification restricts the set of rules to choose from, to only those corresponding to the relations that appear in the graph.
As an example, the document revision pattern depicted in Fig. \ref{figure:wikipedia-pattern} may be expressed as follows, using PROV-N syntax:
\footnote{Domain-specific properties have been added to nodes and relations to denote the role of entities, activities, and agents in the pattern.}
\begin{small}
\ttfamily
\begin{tabbing}
entity(e1, [ prov:type="Document" ])  \\
entity(e2, [ prov:type="Document" ]) \\
activity(a, 2013-11-16T16:00:00, 2013-11-16T16:05:00, [prov:type="edit"]) \\
agent(ag, [ prov:type='prov:Person' ]) \\
used(a, e1, 2013-11-16T16:00:00) \\
wasGeneratedBy(e2, a, -, [ ex:fct="save" ]) \\
wasAssociatedWith(a, ag, -, [ prov:role="contributor" ]) \\
wasDerivedFrom(e2, e1, a)
\end{tabbing} 
\normalfont
\end{small}
\sloppy Using this graph, \provGen{} determines that only $\mathit{wasGeneratedBy}$, $\mathit{used}$, $\mathit{wasDerivedFrom}$ and $\mathit{wasAssociatedWith}$ rules are to be used.
Furthermore, it will associate the properties and values found in the seed graph, for instance \texttt{ex:fct="save"}, to the new nodes.
\paragraph{2. Constraints.} Even with this restriction, unconstrained generation would lead to a graph with arbitrarily high node degree and branching factor, which would bear little resemblance to the seed trace provided, except in its relationship makeup.
To further control the generation process, the second user input consists of an additional set of constraints, specified using a natural and intuitive syntax.

\begin{table}\label{tab:DSL-ex}
\begin{scriptsize}
\ttfamily
\begin{tabularx}{\textwidth}{|X|XX|X|}   \hline
\textbf{Determiner} & \multicolumn{2}{c|}{\textbf{Imperative}} & \textbf{Condition} \\ \hline
& Requirement & Req. qualifier & \\ \hline 
    an Entity & has in degree & at most 1; & \\  
    & & & \\
    an Agent & has relationship "WasAssociatedWith" & between 1, 1000 times, with distribution gamma(..., ...), &
    unless it has relationship "ActedOnBehalfOf"; \\ 
    & & & \\ 
    an Activity & has relationship "Used" & exactly 1 times, & unless it has property \textbraceleft prov\textunderscore type:"create"\textbraceright; \\ 
    &&&\\
    an Entity & has relationship "WasDerivedFrom", & at least 1 times, & unless it has relationship "WasGeneratedBy" with the Activity, a1, AND a1 has property {prov:type="create"};  \\    \hline 
\end{tabularx}
\normalfont
\end{scriptsize}
\caption{Examples of user-defined constraints for graph generation.}
\end{table}

Constraints consist of three structural components, as shown in the examples of Table 1, namely a \textit{determiner}, an \textit{imperative}, and a \textit{condition}.
 The determiner is either variable (\texttt{an Agent}) or invariable (\texttt{the Agent, a1}) and determines the elements to which a constraint applies. 
Requirements on these elements are specified by means of the Imperative clause. 
For instance \texttt{has in degree} (the requirement) \texttt{at most 1} (a qualifier) allows a new incoming edge to be added to any Entity that has none.
The qualifier may optionally include a probability distribution, as in the second example. 
 This determines the likelihood that an action be taken in order to satisfy the requirement, namely 
 the generation of a new $\mathit{WasAssociatedWith}$ relation.
Furthermore, a condition specifies the applicability of an imperative to a determined element, i.e. \texttt{when} (selective condition) or \texttt{unless} (greedy condition).
Thus, the second constraint inhibits the creation of a new $\mathit{WasAssociatedWith}$ relation for any Agent that already has a $\mathit{ActedOnBehalfOf}$ relation associated to it.
Conditions admit the use of logical connectives, as in the third and last constraint examples, and may predicate on properties that are mentioned in the seed graph, such as \texttt{prov:type} (pre-defined) or \texttt{ex:name} (user-defined).
Finally, the last constraint shows an example of variable usage (\texttt{a1}).

\begin{figure}
\centering
\includegraphics[width=.8\textwidth]{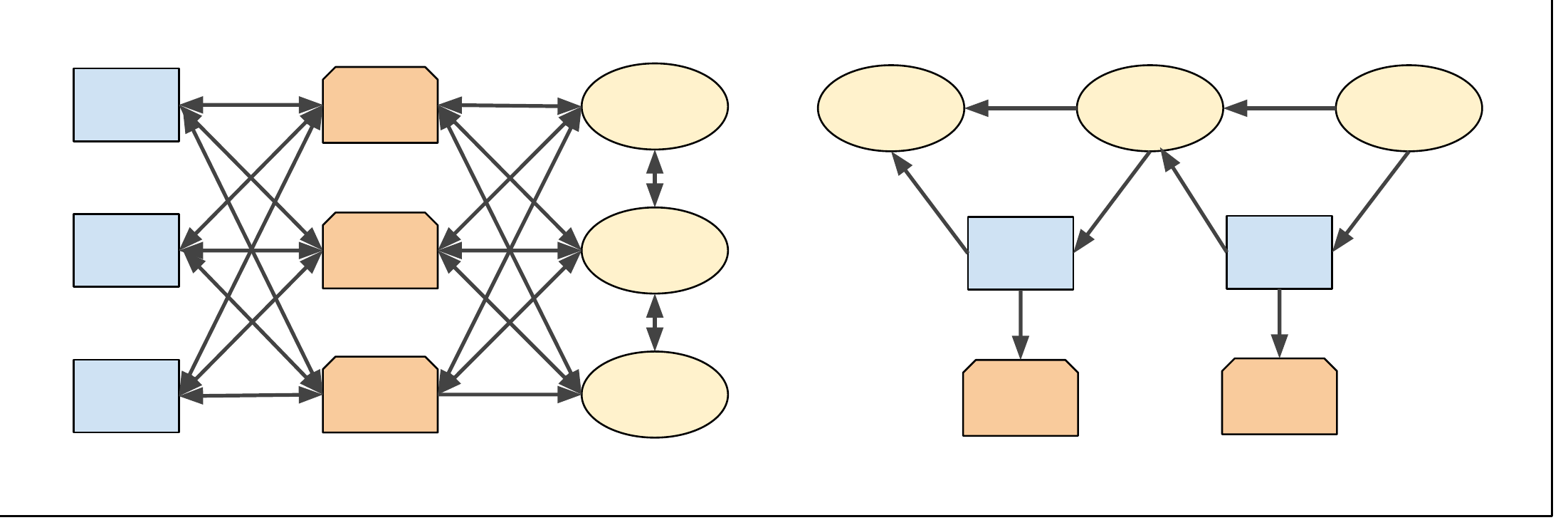}
\caption{Sketch of PROV graphs generated with and without enforcing user constraints}
\label{fig:constraints_noconstraints}
\end{figure}

Note that these constraints are in addition to those defined in the PROV-CONSTR document \cite{w3c-prov-constraints}. 
For instance, \provGen{} will not create a graph where entities are generated by multiple activities.
The sketch in Fig.\ref{fig:constraints_noconstraints} shows the different patterns obtained when generating the graph with and without enforcing the constraints. 

A more complete account of the constraint DSL can be found as part of the \provGen{} documentation to be published shortly.


\paragraph{3. Execution Parameters.} Finally, users may specify additional execution parameters to control the number of distinct (unconnected) graphs to be generated, as well as the average number of nodes and edges per graph.
More advanced parameters can be used to control the average height (maximum depth) and width (maximum breadth) for each graph generated.
The combination of seed graph, constraints, and execution parameters leads to collections of PROV graphs that approximate real traces from different domains, and which can be used to populate selected areas of our provenance state (Fig.\ref{fig:prov-quadrant}).
In Sec. \ref{sec:eval} we briefly sketch the evaluation method we are using to test the quality of generated graphs, with respect to a large testbed of provenance graphs with known topological properties.

Overall, the \provGen{} generation process consists of a nested iteration loop. In the inner loop, \provGen{} iterates over the set of active atomic rewrite rules.
When a rule fires, any constraint that applies to the elements that it is operating upon is checked, and if any of those constraints is violated, the rule has no effect.
This process is repeated in the outer loop, until a halting condition is satisfied, i.e., the desired size is reached, and the DSL constraints are satisfied.

\section{Mapping the model to graph DBMS queries} \label{section:system-architecture}

\provGen{} is implemented using the Neo4J graph DBMS\footnote{The Neo4j project: http://bit.ly/Pwux7U}
 as a back end. 
 In particular, both atomic rewrite rules and user constraints are transparently compiled into \texttt{CREATE} and \texttt{MATCH} statements expressed in Cypher, Neo's declarative graph pattern language\footnote{Cypher documentation: http://bit.ly/1klIlMK}. 
Queries (in addition to \texttt{CREATE} statements) are required at each iteration to test the requirements and conditions associated with user constraints (Table 1).
 This compilation step provides isolation from the data layer, delegating graph traversal to the underlying DBMS, and also provides flexibility for retargeting the graph generator to a different back end.
A native graph DBMS also offers a more natural data model for PROV than a more traditional RDBMS solution.

\begin{figure}
    \includegraphics[width=1.1\textwidth]{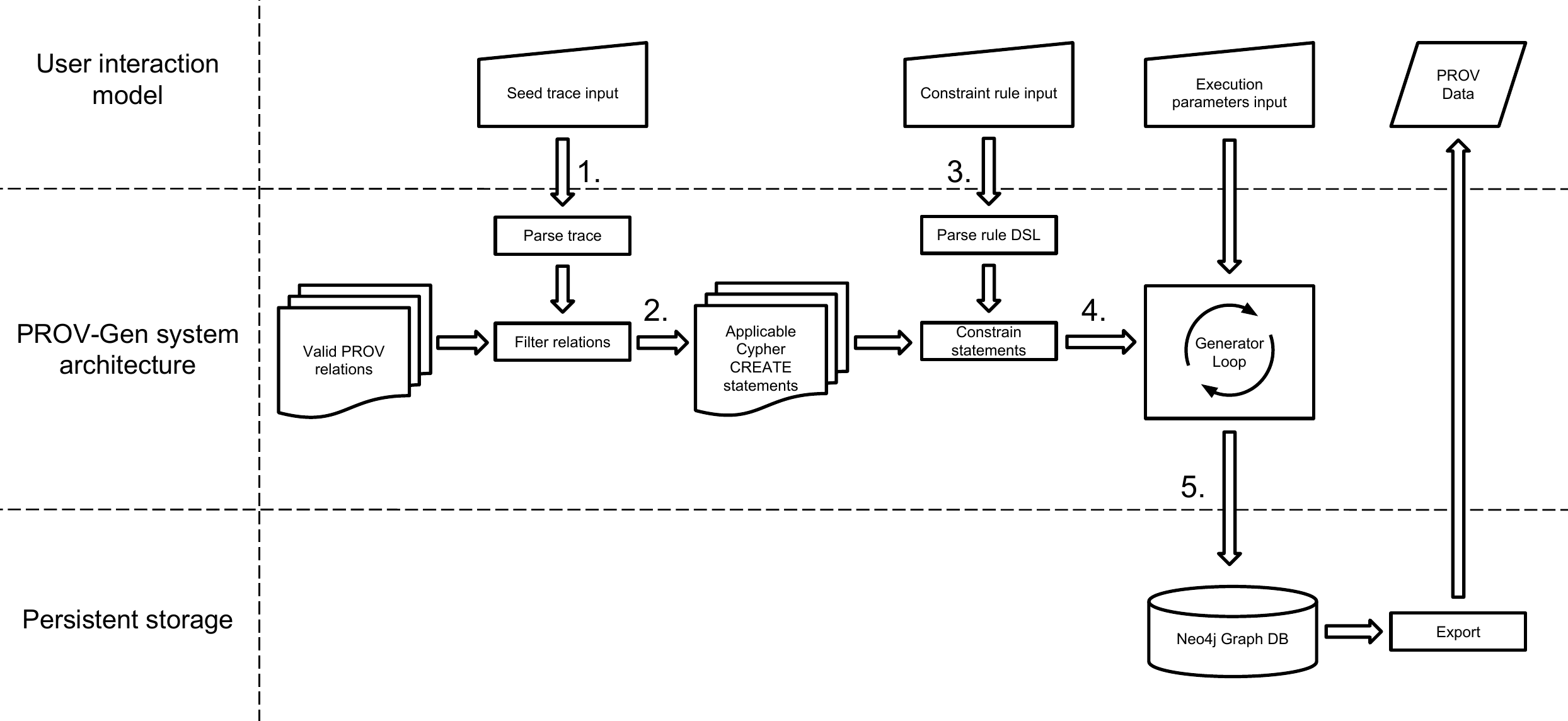}
    \caption{\provGen{} system architecture.}
    \label{figure:architecture-overview}
\end{figure} 
The \provGen{} architecture is shown in Fig. \ref{figure:architecture-overview}.
Components are deployed on a server, which is reachable from a web based client application through a REST API.
In the following sections, we focus on the steps involved in generating Cypher queries from rewrite rules and user constraints.

\subsection{From seed traces to MATCH query clauses}

The first step involves parsing the seed traces. Since these user-supplied samples of PROV data may be serialized into multiple formats, parsing relies upon several third party libraries, including the OWLAPI\footnote{The OWLAPI project: http://bit.ly/N9hsPM}
 and ProvToolbox.\footnote{The ProvToolbox project: http://bit.ly/1fV95nN}
This step results in a subset of the 39 pre-defined atomic graph rewrite rules, mentioned in Sec. \ref{section:prov-gen-model}, to be selected for the generation step.

Rewrite rules are statically mapped to Cypher queries.
As an example, below we show the queries responsible for creating the PROV $\mathit{used}$ relationship. 
Note that multiple queries are required in order to account for the directed nature of PROV relationships and the ability to create a edge between two pre-existing nodes.
\ttfamily
    \begin{small}
            \begin{tabbing}
                (1) \= \textbf{MATCH} (a:Activity \textbraceleft\textbraceright) ~\textbf{CREATE} (a)-[:USED \textbraceleft\textbraceright]-\textgreater(:Entity \textbraceleft\textbraceright) \\
                 (2)\>               \textbf{MATCH} (a:Entity \textbraceleft\textbraceright) ~\textbf{CREATE} (a)\textless-[:USED \textbraceleft\textbraceright]-(:Activity \textbraceleft\textbraceright) \\ 
                (3)\>\textbf{MATCH} (a:Activity \textbraceleft\textbraceright), (b:Entity \textbraceleft\textbraceright) ~\textbf{CREATE} (a)-[:USED \textbraceleft\textbraceright]-\textgreater(b) \\
            \end{tabbing} 
    \end{small}
\normalfont
Query fragment (1) matches any node \texttt{a} of type Activity, it creates a new Entity node, and it connects it to \texttt{a} using a $\used$ relationship. 
Symmetrically, (2) adds a new Activity node to any existing Entity node. 
Finally, (3) takes a pair of \textit{existing} nodes \texttt{a} (Activity), \texttt{b} (Entity) and again creates a $\used$ relationship between them.

The examples above show empty brackets, to indicate that no properties are associated to the nodes and relationships. 
However, all properties associated to the elements of the seed trace are also associated to corresponding elements of the new graph. 
Thus, for example activities would have a property \texttt{prov:type}, inherited from the activity node in the seed graph above.

%
%
%
%
%
%
%

\subsection{Constraints as WHERE clauses}

The DSL parser\footnote{The parser is implemented using Scala parser combinators: http://bit.ly/1cURrAo.} separates the component elements of each constraint, namely 
\textit{determiner}, \textit{imperative} and \textit{condition}.
Requirements may be expressed on various graph features, i.e., nodes in/out degree, relationship, property, etc\ldots. Each type of requirement is compiled into a Cypher query \texttt{WHERE} clause.
These clauses are then added to the \texttt{MATCH} statements that represent the atomic rewrite rules, to form complete queries.
Consider the following example:
\begin{small}
\ttfamily
        \begin{tabbing}
            an Activity \= has relationship ``Used'' exactly 1 times, \\
            \> unless it has property \{``ex:name'':``create''\};\\
            an Activity has degree at most 5;    
        \end{tabbing}
        \normalfont
\end{small}
These constraints are easily interpreted in the context of a document revision pattern, where activities are edits of document versions, which produce a new version. 
For these activities, we stipulate that they use only one entity (the original document). Activities that create new documents are exceptions, noted by the \texttt{ex:name=create} property, and these activities are allowed to use zero or more input documents. 
Additionally, we add an upper bound to an Activity node's degree to illustrate a more complex constraint.

The constraint is compiled into query fragments (4) and (5) in the Cypher query below, where they are merged with the \texttt{MATCH} and \texttt{CREATE} clauses of atomic query (1) from the example above:
\begin{small}
        \begin{tabbing}
            (1) \=\textbf{MATCH} (a:Activity \textbraceleft\textbraceright)\\
            (4) \>\textbf{MATCH} (a)-[r]-() \\
            (5) \> \textbf{WHERE NOT} a.ex\_name = ``create'' \textbf{AND NOT} count(r) $>=$ 5\\
            (1) \>\textbf{CREATE} (a)-[:USED \textbraceleft\textbraceright]-\textgreater(:Entity \textbraceleft\textbraceright)     
        \end{tabbing}
\end{small}
 
The query specifies at the same time the node and relationship generation, and the constraint. 
The \texttt{MATCH} clauses bind variables \texttt{a} and \texttt{r} to an Activity and to the set of its edges, respectively (either incoming or outgoing, as no direction is specified). The \texttt{WHERE} clause ensures that the \texttt{CREATE} statement (which creates a new $\used$ relationship) is only executed on \texttt{a} if the \texttt{ex\_name} property is not ``create'', and the number of edges in set $r$ is at most 4.

\subsection{Generator Loop}~ \label{section:generator-loop}

The generator loop (Fig. \ref{figure:architecture-overview}) accepts a collection of atomic create operations, selected and constrained as described above, and repeatedly iterates over it, executing each associated Cypher query against the underlying graph database.


The generator loop has several halting conditions: both explicit, where execution parameters, detailed in Sec. \ref{section:prov-gen-model}, halt generation as the order $|V|$ and size $|E|$ of the graph reach their specified maxima; and implicit, where constraint rules may prevent the execution of individual operations in order to avoid violating specified range requirements. 
Note that limits in cardinality imposed by execution parameters may be met before the minimum requirements of a constraint rule are satisfied.
When this is the case, \provGen{} gives priority to the user constraints, to ensure that those are not violated. 

%
%

\section{Evaluation methodology}  \label{sec:eval}

%
The main purpose of \provGen{} is to fulfill the need to generate a possibly large number of provenance graphs for data domains where provenance is not yet routinely collected, or is not abundant. 
Yet, our evaluation of the system's effectiveness relies on precisely those domains where large provenance collections are available. 
Specifically, we evaluate \provGen{} by comparing selected properties of existing ``real-world'' provenance graphs, which we call \textit{control} set, to those of generated graphs (the \textit{test} set) intended to emulate them.
Using this approach, we aim to empirically demonstrate that \provGen{} may be configured to generate datasets that are ``similar'' to those produced by multiple different sources of provenance.

Our evaluation is ongoing. Here we use illustrate the approach using one single control set, namely a set of Wikipedia provenance traces, representative of the \textit{document revision} pattern, taken from the ProvBench repository and compliant with PROV.\footnote{\url{http://bit.ly/1fBOswR}}
The control graphs include about 4,000 nodes and 6,000 relationships.
Our test set consists of two synthetic datasets of roughly the same size as the control, produced using \provGen{} with a user-created seed trace for the document revision pattern, along with constraints and parameters. 
 
In this initial evaluation we have considered three simple criteria. 
Firstly, we note that in the control set, which follows the linear Wikpedia pattern (Fig. \ref{figure:wikipedia-pattern}), each Entity is used exactly once.
Thanks to our user constraints, this is easily replicated exactly in the test set.
Secondly, as example criteria we additionally consider the \textit{number of associations per Agent}, and the \textit{average number of entities with distinct titles contributed to, per Agent}. 
In the control, each Agent has 2.4 associations on average (std dev. 6.2), while in our test set it has 2.9.
The average number of contributions per Agent is 1.1 in the control (std dev 0.8), while in the test is 1.8.
Encouraged by these preliminary results, we are now in the process of more extensively testing \provGen{} using a variety of criteria that can be easily measured both on control and on test graph.

\section{Conclusion}

In this paper we have presented \provGen{}, a PROV-specific graph generator driven by user-defined seed graphs, which represent provenance patterns, and additional user-defined constraints designed to enforce semantics properties of the generated graph.
Constraints are expressed in a dedicated ``plain english'' constraints language.

One feature that sets \provGen{} apart from existing approaches to graph generation is that it provides users with local control over topological features and statistical characteristics of the graph.
Constraints are evaluated locally for each node created, thus avoiding the complexity of verifying them globally.
\provGen{} is implemented using a Neo4J graph database back end. Graph rewrite rules and user constraints are both mapped to Cypher queries. Rewrite rules are mapped to \texttt{CREATE} clauses, while constraints are compiled into \texttt{WHERE} clauses. The two are blended together into complete Cypher queries, so that graph generation relies entirely on Neo4J's native query engine.

We have also briefly discussed our approach to evaluating the effectiveness of \provGen{} in generating ``real-world'' provenance, i.e., by comparing some of its key statistical properties with those of real graphs within the same class.
We are currently experimenting with a variety of seed graph patterns, and more extensively evaluating \provGen's capability to mimick real provenance. 

Graph generation performance is another concern we are currently addressing. Generating large scale graphs requires efficient execution of the \texttt{MATCH}--\texttt{CREATE}-\texttt{WHERE} queries shown above, on graphs of increasing size.
We are finding that Neo4J may not be an optimal choice, as it is geared for OLTP workloads with consequent transaction management overhead. 
However, our architecture is flexible and allows for experimentation, as changing the back end simply requires retargeting the mapping of rules and constraints to a different query language.

\bibliographystyle{llncs2e/splncs}
\bibliography{prov_storage_and_analysis-IPAW14}

\end{document}